\documentclass[twocolumn, nofootinbib, superscriptaddress, floatfix, longbibliography]{revtex4-1} 

\usepackage{multirow}
\usepackage{booktabs}
\usepackage{placeins}
\usepackage{soul}
\usepackage{bbm}
\usepackage{amsmath,amssymb,amsbsy,bm,amsfonts,mathrsfs}
\usepackage{graphicx}
\usepackage{wrapfig}
\usepackage{url}
\usepackage{float}
\usepackage{lipsum}
\usepackage{enumitem}
\usepackage{titlesec}
\usepackage{microtype}
\usepackage{slashed}
\usepackage[normalem]{ulem}
\usepackage{braket}

\usepackage[usenames,dvipsnames,svgnames]{xcolor}
\definecolor{redd}{rgb}{0.8, 0.1,0.2}
\definecolor{navy}{rgb}{0.05, 0.23,0.75}

\usepackage{color}

\definecolor{Orange}{cmyk}{0,0.61,0.87,0}
\definecolor{JungleGreen}{cmyk}{0.99,0,0.52,0}
\definecolor{OliveGreen}{cmyk}{0.64,0,0.95,0.40}
\definecolor{Brown}{cmyk}{0,0.81,1,0.60}
\definecolor{RoyalBlue}{cmyk}{0.71,0.53,0,0.12}
\definecolor{Gray}{cmyk}{0,0,0,0.40}
\definecolor{LightPink}{cmyk}{0.0,0.25,0,0}
\definecolor{LLightPink}{cmyk}{0.0,0.10,0,0}
\definecolor{LightBlue}{cmyk}{0.25,0,0,0}
\definecolor{LightGray}{cmyk}{0,0,0,0.2}

\usepackage{xcolor}
\definecolor{gesfpurple}{rgb}{0.47,0.19,0.42}

\definecolor{gesflanse}{rgb}{0.00,0.50,0.50}

\definecolor{gesfblue}{rgb}{0.08,0.42,0.76}

\definecolor{gesfred}{rgb}{1,0,0}

\definecolor{gesfwhite}{rgb}{1,1,1}

\definecolor{gesfblack}{rgb}{0,0,0}

\newcommand{\geqn}[1]{Eq.\,\hypersetup{linkcolor=blue}(\ref{#1})\hypersetup{linkcolor=blue}}
\newcommand{\gfig}[1]{{\hypersetup{linkcolor=violet}Fig.\,\ref{#1}\hypersetup{linkcolor=blue}}}

\usepackage[colorlinks]{hyperref}
\hypersetup{
    colorlinks    =    true,
    citecolor     =    navy,
	linkcolor     =    redd,
	urlcolor      =    navy,
	anchorcolor   =    blue
}

\makeatletter
\def\bibsection{%
   \par
   \begingroup
    \baselineskip26\p@\bib@device{\hsize}{72\p@}%
   \endgroup
   \nobreak\@nobreaktrue
   \addvspace{19\p@}%
  }%
\makeatother

\graphicspath{{figs/}}
\allowdisplaybreaks[4]

\begin{document}

\title{Nuclear Production and Analytic Attenuation of Energetic MeV Solar Dark Matter}

\author{Shao-Feng Ge}
\email[Corresponding Author: ]{gesf@sjtu.edu.cn}
\affiliation{State Key Laboratory of Dark Matter Physics, Tsung-Dao Lee Institute \& School of Physics and Astronomy, Shanghai Jiao Tong University, Shanghai 200240, China}
\affiliation{Key Laboratory for Particle Astrophysics and Cosmology (MOE) \& Shanghai Key Laboratory for Particle Physics and Cosmology, Shanghai Jiao Tong University, Shanghai 200240, China}

\author{Jie Sheng}
\email{shengjie04@sjtu.edu.cn}
\affiliation{State Key Laboratory of Dark Matter Physics, Tsung-Dao Lee Institute \& School of Physics and Astronomy, Shanghai Jiao Tong University, Shanghai 200240, China}
\affiliation{Key Laboratory for Particle Astrophysics and Cosmology (MOE) \& Shanghai Key Laboratory for Particle Physics and Cosmology, Shanghai Jiao Tong University, Shanghai 200240, China}

\author{Chen Xia}
\email{xiac@sari.ac.cn}
\affiliation{State Key Laboratory of Dark Matter Physics, Tsung-Dao Lee Institute \& School of Physics and Astronomy, Shanghai Jiao Tong University, Shanghai 200240, China}
\affiliation{Key Laboratory for Particle Astrophysics and Cosmology (MOE) \& Shanghai Key Laboratory for Particle Physics and Cosmology, Shanghai Jiao Tong University, Shanghai 200240, China}
\affiliation{Shanghai Synchrotron Radiation Facility, Shanghai
Advanced Research Institute, Chinese Academy of Sciences,
Shanghai 201204, China}

\author{Chuan-Yang Xing}
\email[Corresponding Author: ]{cyxing@upc.edu.cn}
\affiliation{State Key Laboratory of Dark Matter Physics, Tsung-Dao Lee Institute \& School of Physics and Astronomy, Shanghai Jiao Tong University, Shanghai 200240, China}
\affiliation{Key Laboratory for Particle Astrophysics and Cosmology (MOE) \& Shanghai Key Laboratory for Particle Physics and Cosmology, Shanghai Jiao Tong University, Shanghai 200240, China}
\affiliation{College of Science, China University of Petroleum (East China), Qingdao 266580, China}

\begin{abstract}
We propose a solar production mechanism of MeV
dark matter to overcome the energy
threshold in direct detection experiments.
In particular, the proton and deuteron fusion
to ${}^3 \mathrm{He}$ of the $pp$ chain that
produces energetic neutrino and gamma photon
with 5.5\,MeV of energy release
can also produce a pair of dark matter particles.
Besides, we establish an analytical formalism of using the
Boltzmann equation to study the solar attenuation
effect on the produced dark matter flux. The projected
sensitivity is illustrated with Argon target at
the DarkSide-LowMass experiment.
\end{abstract}

\maketitle

\section{Introduction} \label{sec:intro}

There are ample evidences of the existence of dark
matter (DM) from cosmological and astrophysical
observations
\cite{Young:2016ala,Bauer:2017qwy,Lin:2019uvt,Arbey:2021gdg}.
The current direct detection experiments are sensitive
to DM with mass $\gtrsim \mathcal O({\rm GeV})$
\cite{Cooley:2021rws}. 
In particular, the most stringent sensitivity on the spin-independent DM-nucleon scattering cross section $\sigma_\mathrm{SI}$ reaches $\mathcal{O}( 10^{-47} ) \, \mathrm{cm}^2$ \cite{PandaX-4T:2021bab,LZ:2022lsv,XENON:2023cxc}. 
On the other hand, the sub-GeV mass range is much less
constrained with not enough energy to overcome the
recoil energy threshold.

Various new detection approaches with lower detection
thresholds have been proposed to increase the
sensitivity for light DM, including the Bremsstrahlung
\cite{Kouvaris:2016afs} and Migdal
\cite{Ibe:2017yqa,Dolan:2017xbu,Baxter:2019pnz,Essig:2019xkx}
effects,
fermionic absorption \cite{Dror:2019onn, Dror:2019dib, Dror:2020czw, Ge:2022ius, Li:2022kca, Ge:2024euk} and nucleon consumption
\cite{Ema:2024wqr, Ge:2024lzy} scenarios,
as well as new detection materials
\cite{Hochberg:2015pha,Schutz:2016tid,Hochberg:2019cyy,Caputo:2020sys,Esposito:2022bnu}.
In addition, new sources of energetic DM can also help to overcome the detection threshold.
For example, DM can be boosted by semi-annihilation
\cite{Berger:2014sqa,Toma:2021vlw}, cosmic rays
\cite{Yin:2018yjn,Cappiello:2018hsu,Bringmann:2018cvk,Ema:2018bih,Cappiello:2019qsw,Dent:2019krz,Bondarenko:2019vrb,Wang:2019jtk,Guo:2020drq,Ge:2020yuf,Cao:2020bwd,Lei:2020mii,Xia:2020apm,Feng:2021hyz,Xia:2021vbz,Xia:2022tid,Wang:2023wrx,Lu:2023aar}, blazars
\cite{Wang:2021jic,Granelli:2022ysi}, the nearest active galactic nucleus Centaurus A
\cite{Xia:2024ryt}, cosmic and supernova neutrinos
\cite{Jho:2021rmn,Das:2021lcr,Chao:2021orr,Lin:2022dbl,Das:2024ghw}, solar reflection
\cite{An:2017ojc,Emken:2021lgc,An:2021qdl}, etc.
Besides boosting the existing DM particles, their decay \cite{Kopp:2015bfa,Bhattacharya:2016tma} or annihilation \cite{Agashe:2014yua} can produce relativistic dark particles.
In addition, boosted dark particles can also evaporate from black holes \cite{Calabrese:2021src}
or appear in the cosmic ray dump in the Earth atmosphere \cite{Alvey:2019zaa,Su:2020zny,Arguelles:2022fqq,PandaX:2023tfq}.

Those boosting mechanisms are all related to
astrophysical or atmospheric processes or origins.
Of them, the solar reflection with acceleration
by thermal electrons inside Sun can only be measured
by the electron recoil signal. Corresponding to a typical
temperature around $15$ million kelvins, the energy is in the
$\mathcal O({\rm keV})$ range which is still far from
overcoming the detection threshold with nuclei recoil.
However, the nuclear fusion inside Sun is intrinsically
at the $\mathcal O({\rm MeV})$ scale. The corresponding
energy release is large enough to produce energetic
DM and subsequently nuclei recoil above the detection
threshold.

\begin{figure}[t]
\centering
\includegraphics[width=8cm]{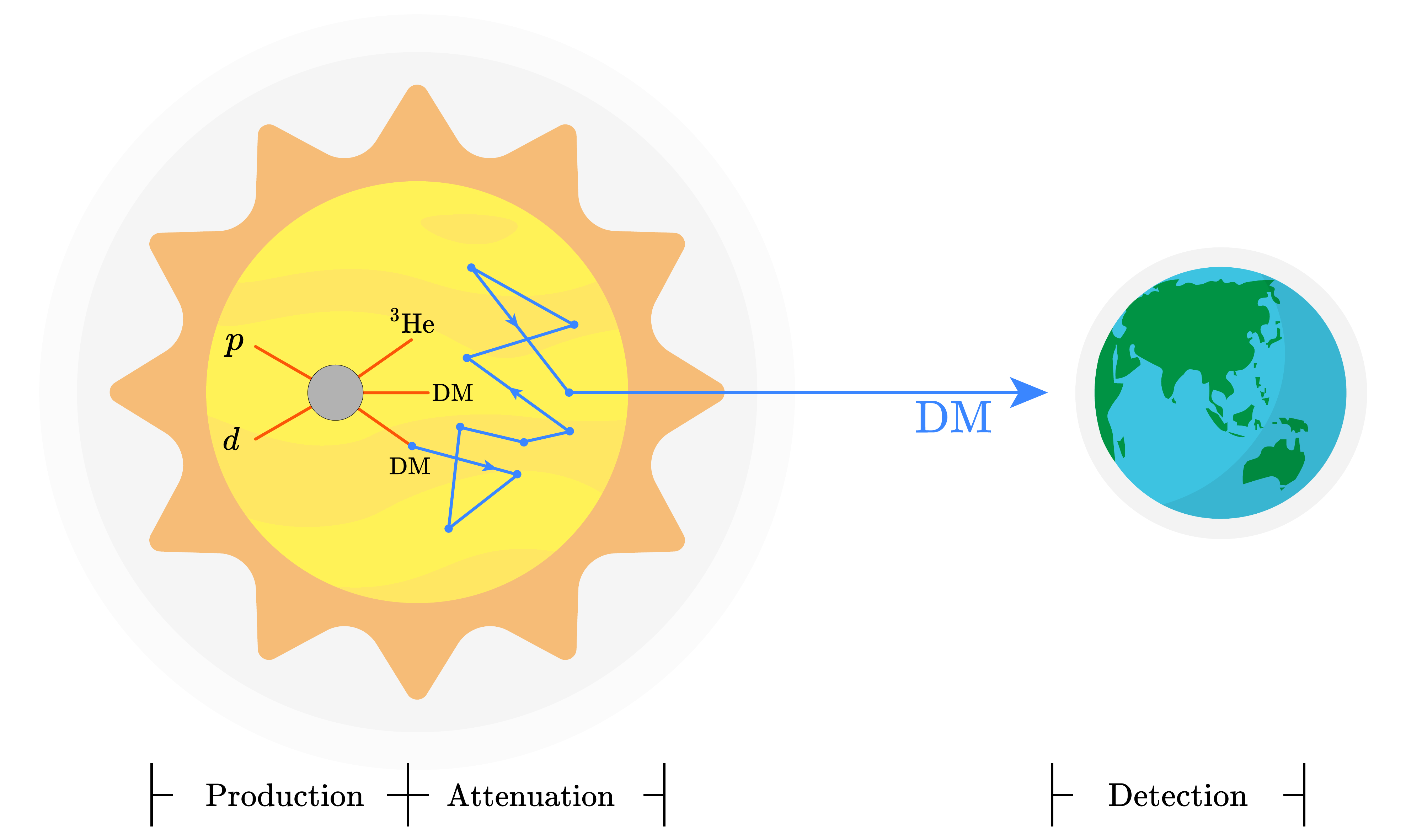}
\caption{Schematic illustration of the production, attenuation and detection of energetic solar DM.}
\label{fig:illustration}
\end{figure}

We propose a possible way of producing energetic MeV DM
from the solar $pp$ chain to overcome the direct
detection threshold with nuclei recoil.
\gfig{fig:illustration} sketches the three key processes,
1) the production of MeV DM from the solar $pp$ chain,
2) the DM scattering with nuclei and the resultant solar
attenuation,
3) the DM direct detection on our Earth,
to be elaborated below.

\section{MeV Solar Dark Matter Production from $pp$ Chain}
\label{sec:production}

Although the solar nuclear fusion process contains both $pp$
chain and CNO cycle \cite{Adelberger:2010qa},
the latter contributes only 1\% to the
energy production and hence can be ignored.
There are three
photon emission processes in the $pp$ chain. Of them,
$p + {}^7 \mathrm{Be} \to {}^8 \mathrm{B} + \gamma$
contributes less than 1\%. Though
${}^3 \mathrm{He} + {}^4 \mathrm{He} \to {}^7 \mathrm{Be} + \gamma$
has a sizable branching ratio of 16.7\%, the released energy
of $1.6 \, \mathrm{MeV}$ is not enough to overcome the
nuclear recoil detection threshold. Only the fusion of proton
($p$) with deuteron ($d$), $p + d \to {}^3 \mathrm{He} + \gamma$, 
has large enough branching ratio (100\%) and energy release
at $5.5 \, \mathrm{MeV}$ \cite{Adelberger:2010qa}. So the MeV
DM production is mainly through $p+d \rightarrow {}^3{\rm He} + X$
where $X$ denotes a group of DM particles.
Due to stability, the DM particle usually appears in
pair,
$p + d \to {}^3 \mathrm{He} + \chi^* + \chi$,
with $X \equiv \chi^* \chi$.
We dub such DM as
\textit{Solar Dark Matter}, in the same sense as
solar neutrino \cite{Bergstrom:2016cbh,Vinyoles:2016djt} or solar axion \cite{CAST:2009klq,Borexino:2012guz,Bhusal:2020bvx,Vergados:2021ejk}.

The momentum transfer, same order as the 5.5\,MeV released energy, corresponds to a length
of 35\,fm which is larger than the size of $p$ and $d$. So one
may neglect their internal structures. In addition, the DM
production involves the scattering of $p$ and $d$
initial states into a bound state ${}^3{\rm He}$ under the
influence of Coulomb and nuclear potentials \cite{Bertulani2019aaa}.
While the initial state is taken as an ionized state of the $p$-$d$
system, the product ${}^3 \mathrm{He}$ is the ground state. 
The fusion process can then be viewed as a transition from
the ionized state to the ground state \cite{Bertulani:2003kr,Huang:2008ye} by
emitting a DM pair.

{\bf Matrix Element} -- We consider a simple interaction of a complex scalar DM $\chi$
and proton for illustration,
$\mathcal{L}
\equiv \frac{1}{\Lambda} \chi^* \chi \bar{p} p$, 
where $\Lambda$ is a cutoff scale.
Using its non-relativistic form \cite{Weinberg:1995mt,Ge:2021snv} for the
DM coupling with the fermionic
$p$-$d$ system, the fusion matrix element reads,
    \begin{align}
        T 
        &= 
        \left\langle f; \bm{p}_{\chi^*}, \bm{p}_\chi \right| 
        \int d^3 \bm{x} dt
        \frac{i}{\Lambda}
        \left[  
            \chi^*(\bm{x}_{p}, t) \chi(\bm{x}_{p}, t) 
            \psi^\dagger(\bm{x},t) \psi(\bm{x},t) 
        \right. \nonumber \\
        &\hspace{1.5cm}
        \left.
            + 
            \chi^*(\bm{x}_{d}, t) \chi(\bm{x}_{d}, t) 
            \psi^\dagger(\bm{x},t)  \psi(\bm{x},t) 
        \right] 
        \left| i \right\rangle , 
    \label{matrix_element}
    \end{align}
where $
    \psi( \bm{x}, t ) 
    = 
    \sum_n \hat{a}^{pd}_n \phi_n (\bm{x}) 
    e^{-i E_n t}
    + 
    \mathrm{h.c.} 
$ is the second-quantized field for the $p$-$d$
system \cite{Ge:2021snv}.
The energy eigenvalue $E_n$ and wave function $\phi_n$ are
solved by the Schrödinger equation with the potential in
\geqn{pd_potential} while $\hat{a}^{pd}_n$ is the annihilation
operator for the corresponding state.
The two terms stand for the contributions from proton at
${\bm x}_p$ and deuteron at $\bm{x}_d$, respectively. 
In the center-of-mass frame, $\bm{x}_p \simeq \frac{2}{3} \bm{x}$ and $\bm{x}_d \simeq -\frac{1}{3} \bm{x}$.
Therefore, from $T \equiv (2\pi) \delta(E_\chi + E_{\chi^*} + E_f - E_i)    \mathcal{M}$, we can extract the scattering amplitude,
\begin{equation}
    \mathcal{M}
    =
    \frac{i}{\Lambda} 
    \int d^3 \bm{x} 
    ( e^{-i \frac{2}{3} \bm{q} \cdot \bm{x}} + e^{i \frac{1}{3} \bm{q} \cdot \bm{x}}) 
    \phi^\dagger_f(\bm{x})  \phi_i(\bm{x}),
    \label{DM_production_amplitude}
\end{equation}
where the momentum transfer $\bm{q} \equiv \bm{p}_\chi + \bm{p}_{\chi^*}$ is the total DM momentum.
The momentum transfer is smaller than the energy release
(which is approximately the size of binding energy
$|E_b| = 5.5$\,MeV)
since $|{\bm q}| \leq |{\bm p}_\chi| + |{\bm p}_{\chi *}|$
and $E_\chi + E_{\chi *} \simeq |E_b|$.
Namely, $|\bm{q}| \lesssim |E_b| = 5.5 \, \mathrm{MeV}$,
and consequently the bound-state wave function is predominantly localized within the region $|\bm{x}| \lesssim 
1/\sqrt{2 m_p |E_b|} \simeq (100 \, \mathrm{MeV})^{-1}$
where $m_p$ is the proton mass.
Thus, $\bm{q} \cdot \bm{x} \lesssim 1/20$
is a small quantity for Taylor expansion.
The leading order of the amplitude vanishes since the 
initial-state wave function $\phi_i(\bm{x})$ and
its final-state counterpart $\phi_f(\bm{x})$
are orthogonal to each other. The nonzero contribution
then appears at the linear order as an E1 transition,
\begin{equation}
  \mathcal{M}
\simeq
  \frac 1 {3 \Lambda} 
  \int d^3 \bm{x} 
  \left( \bm{q} \cdot \bm{x} \right)
  \phi^\dagger_f(\bm{x})  \phi_i(\bm{x})
\equiv
  \frac 1 {3 \Lambda} \bm q \cdot \langle f | \bm x | i \rangle,
\label{DM_production_amplitude_linear_term}
\end{equation}
where $\langle f | \bm x | i \rangle \equiv \int
\bm x\phi^\dagger_f(\bm x) \phi_i(\bm x) d^3 \bm x$
captures the dipole structure.

{\bf Differential Cross Section}
-- With the matrix element \geqn{DM_production_amplitude_linear_term} above, the transition rate $d \Gamma_{fi}$ between the initial and final states can be obtained via the
Fermi's Golden Rule \cite{Sakurai:2011zz},
\begin{equation}
  d\Gamma_{fi}
=
  \overline{|\mathcal{M}|^2} (2\pi) \delta(E_i - E_f - E_\chi - E_{\chi^*})
  d \Phi_n,
\label{eq:dwfi}
\end{equation}
where $d \Phi_n$ is the $n$-body final-state phase space,
\begin{equation}
  d \Phi_n
\equiv 
  \frac {d^3 \bm{p}_\chi} {(2\pi)^3 2 E_\chi}
  \frac{d^3 \bm{p}_{\chi^*}}{(2\pi)^3 2 E_{\chi^*}}.
\end{equation}
Note that it only contains the DM phase space.
Since the final bound state $f$ is fixed, as
will be shown later, there is no continuous
degree of freedom in its quantum state.
Taking the matrix element in our case
\geqn{DM_production_amplitude_linear_term}
into \geqn{eq:dwfi}, one gets,
\begin{align}
    d &\Gamma_{fi}
     = 
    \left(
      \frac{1}{6} \frac{1}{9 \Lambda^2} 
      \sum_{\mathrm{spins}} 
      \left| \bm{q} \cdot \langle f | \bm{x} | i \rangle \right|^2 
    \right) (2\pi) 
  \label{transition_rate_DM_1}
    \\
    & \times 
    \delta(E_i - E_f - E_\chi - E_{\chi^*})
    \frac{d^3 \bm{p}_\chi}{(2\pi)^3 2E_\chi} 
    \frac{d^3 \bm{p}_{\chi^*}}{(2\pi)^3 2E_{\chi^*}}, 
    \nonumber
\end{align}
where the prefactor $1/6$ comes from the average
over the initial proton with spin $1/2$ and deuteron
with spin $1$.
Through the relationship between the momentum transfer and the momentum of the final-state DM particles
${\bm q} = {\bm p}_\chi + {\bm p}_{\chi^*}$, the integration over ${\bm p}_{\chi^*}$ can be transformed into the integration over ${\bm q}$.
This would allow factorization of
$\bm q$ and $\langle f |\bm x| i \rangle$ by the
zenith angle ($d \cos \theta_q$) integration,
$\int |\bm q \cdot \langle f |\bm x| i \rangle|^2 d \cos \theta_q = 2 |\bm q|^2 |\langle f |\bm x| i \rangle|^2$/3.
Then the $\bm q$ integration contains a
$\int |{\bm q}|^4 d|{\bm q}|$. By using the energy conservation delta function, the integration over the angle can yield the factor $E_{\chi^*}/|{\bm q}| |{\bm p}_\chi|$ since $E_{\chi^*} \equiv \sqrt{|{\bm q}|^2 - 2 |{\bm q}| |{\bm p}_{\chi}| \cos \theta_{|{\bm q}|, |{\bm p}_{\chi}|} + E_\chi^2}$. Taking this into consideration, the integration contains $\int |{\bm q}|^3 d|{\bm q}|$, and the transition rate should be proportional to $|{\bm q}|^4$ as,
\begin{equation}
    \frac{d \sigma_\chi  }{dE_\chi}
    =
    \frac{|\bm{q}|_{\max}^4-|\bm{q}|_{\min}^4}{10368 \pi^3 v_{pd} \Lambda^2}
    \sum_\mathrm{spins} 
    |\langle f | \bm x | i \rangle|^2 ,
    \label{dsigmadE}
\end{equation} 
where the maximal and minimal momentum transfer are defined as $|\bm{q}|_{\max,\min} \equiv \left| |\bm{p}_\chi| \pm |\bm{p}_{\chi^*}| \right|$, and the summation is taken over all the initial and final particle spins. 
The factor $v_{pd}$ in the denominator is the asymptotic
relative velocity between proton and deuteron.
The above differential cross section is the transition rate $d \Gamma_{fi}$ over a flux factor $\mathcal F$. 
Due to the interaction between the proton and deuteron, the incident wave function of the system in the center-of-mass frame is no longer a plane wave, but has the form of 
$\phi_i({\bm x}) \sim e^{i{\bm k} \cdot {\bm x}} + f(\theta) e^{i k |{\bm x}|}/|{\bm x}| $
\cite{sobelman2016introduction}, whose specific solution will be discussed in detail in the next subsection. According to the definition of the flux factor in the polar coordinate system, the flux factor is 
$\mathcal F \equiv v_{pd} \times \left( |\phi_i({\bm x})|^2 |{\bm x}|^2 d \Omega/ |{\bm x}|^2 d\Omega \right) = v_{pd}$ at $|{\bm x}| \to \infty$
where the plane wave component $e^{i \bm k \cdot \bm x}$ dominates.

{\bf Wave Functions}
-- 
In order to calculate the differential cross section \geqn{dsigmadE} and later the DM production rate, one needs
to specify the initial- and final-state wave functions
$\phi_i$ and $\phi_f$.
The potential of the $p$-$d$ system contains three parts
\cite{Bertulani2019aaa},
\begin{equation}
    V(\bm x)
=
  V_0(x)
+ V_S (x) (\bm l \cdot \bm s_p)
+ V_C(x),
\label{pd_potential}
\end{equation}
where $\bm x \equiv \bm x_p - \bm x_d$ is the
relative distance between proton ($\bm x_p$) and deuteron
($\bm x_d$) with $x \equiv |\bm x|$. The nuclear
and spin-orbital interactions $V_0(x)$ and $V_S(x)$
are parametrized as
\begin{equation}
  V_0 (x) = V_0 f_0(x),
\quad
  V_S (x) = -V_{S} 
  \frac{1}{m_\pi^2} 
    \frac{1}{x} \frac{df_S(x)}{dx},
    \label{eq:short-int}
\end{equation}
using the Woods-Saxon potential with pion mass $m_\pi$
and~\cite{Woods:1954zz}
\begin{equation}
  f_{0,S}(r) = \left[ 1 + \exp\left( \frac{r-R_{0,S}}{a_{0,S}} \right) \right]^{-1} .
\end{equation}
For the process of $p$-$d$ fusion, the parameterization constants are~\cite{Huang:2008ye},
\begin{subequations}
\begin{align}
    V_0 &= -44.43 \,\mathrm{MeV}, & R_0 &= R_{S} = 1.803 \,\mathrm{fm},  \\
    V_{S} &= -10 \,\mathrm{MeV}, & a_0 &= a_{S} = 0.65 \,\mathrm{fm}.
\end{align}
\end{subequations}
The spin-orbital term contains the orbital angular momentum operator
$\bm l$ and the proton spin operator $\bm s_p$.
Finally, $V_C(x)$ is the Coulomb potential,
\begin{equation}
  V_C(x) = 
  \begin{cases}
    \frac{Z_p Z_d e^2}{x} & x>R_C, \\
    \frac{Z_p Z_d e^2}{2R_C} \left( 3-\frac{x^2}{R_C^2} \right) & x< R_C.
  \end{cases}
\end{equation}
Here, $Z_{p\,(d)} = 1$ denotes the charge number of
$p\,(d)$ and $R_C = 1.803 \,\mathrm{fm}$ is the charge
radius \cite{Huang:2008ye}.

Since the typical energies are $\mathcal O({\rm MeV})$ at most,
the initial ionized state wave function $\phi_i (\bm{x})$
and its final bound state counterpart $\phi_f (\bm{x})$ of
the $p$-$d$ system are solved with the
non-relativistic Schrödinger equation.
Owing to the spherical symmetry of the interacting potential \geqn{pd_potential}, the wave function of any state can be decomposed into a radial and an angular component as,
\begin{equation}
  \Psi_{EJM} (\bm{x})
\equiv
  \frac{u^J_{l j}(E,x)}{x} \mathcal{Y}^{lj}_{JM}.
\label{wave_function_decomposition}
\end{equation}
where $E$ is the energy eigenvalue of the state.
Since the $p$-$d$ system contains two spin information
($s_p$, $m_{s_p}$) and ($s_d$, $m_{s_d}$) for
the proton and deuteron, respectively, we choose to
first couple the orbital angular momentum wave
function $Y_{l,m_{l}}$ with the proton spin
wave function $\left| s_p m_{s_p} \right\rangle$ to give
$ \left| j m_{j} \right\rangle \equiv
\sum \limits_{m_{l}, m_{s_p}} 
\left\langle l m_{l} s_p m_{s_p} | j m_{j} \right\rangle
Y_{l,m_{l}} \left| s_p m_{s_p} \right\rangle $. 
Then $\left| j m_{j} \right\rangle$ further couples
with the deuteron spin wave function
$\left| s_d m_{s_d} \right\rangle$ to give
the angular wave function,
\begin{equation}
  \mathcal{Y}^{lj}_{JM} \equiv \sum_{m_{j},m_{s_d}} \left\langle j m_{j} s_d m_{s_d} | J M \right\rangle \left| j m_{j} \right\rangle \left| s_d m_{s_d} \right\rangle.
\end{equation}
Note that the proton is spin half ($s_p = 1/2$)
while deuteron has spin one ($s_d = 1$). The factors
labeled as $\left\langle \cdot | \cdot \right\rangle$
denotes the Clebsch–Gordan coefficients.

The radial wave function $u^J_{l j}$ is the solution of the radial Schrödinger equation,
\begin{equation}
  \left[ - \frac{\hbar^2}{2 \mu_{pd}} \left( \frac{d^2}{dx^2} - \frac{l(l+1)}{x^2} \right)  + V(x) 
  \right] u^J_{l j} = E\, u^J_{l j},
  \label{radial_Schrödinger_equation}
\end{equation}
where 
the inner product of angular momenta in \geqn{pd_potential} is $\boldsymbol{\mathrm{l}} \cdot \boldsymbol{\mathrm{s}}_p \equiv \left[j(j+1)-l(l+1)-s_p(s_p+1)\right]/2$ and 
the reduced mass is $\mu_{pd} \equiv m_p m_d /(m_p+m_d)$.

The initial ionized state with
center-of-mass energy $E$
can be described by a superposition of the eigen-wave-functions in \geqn{wave_function_decomposition} with different angular momenta.
We can define a parameter called the asymptotic momentum $k \equiv \sqrt{2 \mu_{pd} E}$ and label such a state as, 
\begin{equation}
  |i\rangle
\equiv
  \sum_{l j J M} C_{l j J M}^{ k m_{s_p} m_{s_d} }  |k, [ (l s_p) j s_d ] J M \rangle.
  \label{initial_state}
\end{equation} 
Here, the structure of the brackets represents the
order of angular momentum coupling. The quantum numbers
$l$ and $s_p$ within the parentheses are first coupled to form 
$j$ in the square brackets, which further couples with $s_d$
to yield the total angular momentum $J$.
Along the incident direction, the plane wave is
expanded as $e^{i|{\bm k}||{\bm r}| \cos \theta}
\equiv \sum_l (2 l+ 1) i^l j_l (|{\bm k}||{\bm r}|)
P_l (\cos \theta)$ without dependence on the
azimuthal angle. Since the Legendre function
$P_l$ is proportional to the spherical harmonic
function $Y_l^0$ with magnetic quantum number
$m = 0$, the expansion coefficients are, 
\begin{align}
  C_{l j J M}^{ k m_{s_p} m_{s_d} }  &\equiv   \frac{\sqrt{(4\pi)(2l+1)}}{k} i^l 
\nonumber  \\
  &\times
  \braket{ j m^{}_{s_p} | l 0 s_pm^{}_{s_p} } 
  \braket{ JM | j m^{}_{s_p} s_d m^{}_{s_d} } .
\end{align}

Both the nuclear and spin-orbital interactions $V_0$ and $V_S$ are short-range, if we consider the region that $x$ is large enough, only the Coulomb potential survives. 
Therefore, the initial ionized radial wave function $u_{lj}^J(k, x \to \infty)$ has a general form of solution as,
\begin{equation}
  u_{lj}^J= 
  \frac{i}{2}
  \left[ H_l^{(-)}(\eta, kx)-S_l H_l^{(+)}(\eta, kx)\right]e^{i \sigma_l} 
  \label{Coulomb_plus_finite_range_radial_WF}
\end{equation}
with $H_l^{(\pm)}$ being the spherical Coulomb functions~\cite{Gaspard:2018xgb}.
Here, the parameters are defined as $\eta \equiv Z_p Z_d \mu_{pd}/4\pi k$ and $\sigma_l \equiv \arg \Gamma (1+l)$. 
The influence of short-range interactions shows up in the factor
$S_l \equiv e^{2 i \delta_l}$ where $\delta_l$ is the corresponding phase shift and can be determined by matching \geqn{Coulomb_plus_finite_range_radial_WF} with the Schrödinger equation solution 
\geqn{radial_Schrödinger_equation} at small radius.
The wave function \geqn{Coulomb_plus_finite_range_radial_WF} is valid only at larger radii (e.g., $x > x_0$ where $x_0 \gtrsim 10^{-15}\,$m). For $x < x_0$, the wave function $u_s$ must be obtained by numerically solving the Schrödinger equation with the boundary condition $u_s (x = 0) = 0$ \cite{cohen1977quantum} and an unknown derivative condition $u'_s (x = 0) = y$. At $x = x_0$, 
the continuity condition requires that the two wave
functions match both in value and in their first derivatives.
This allows determination of the derivative $y$ and the phase shift $\delta_l$.

The final state is the ground state of the bound $p$-$d$
system with energy level $n = 0$, since ${}^3$He has no
other excited states \cite{IAEALiveChart}.
It also has the fixed ground-state angular quantum numbers
as, $l = 0$, $j = \frac 1 2$, and $J = \frac 1 2$.
As a result, the final state is simply,
\begin{equation}
  |f\rangle = 
  \left|n=0, \left[ \left( 0 \frac{1}{2} \right) \frac{1}{2} 1 \right] \frac{1}{2} M'  \right\rangle ,
  \label{final_state}
\end{equation}
where $M'$ denotes the spin direction of the ${}^3 \mathrm{He}$ state.
For a bound state, the short-range interactions in
\geqn{eq:short-int} cannot be neglected. 
Consequently, the corresponding radial wave function
must resort to numerical solutions and it should 
satisfy the boundary conditions that
$u^J_{l j} (x=0) = u^J_{l j}(x=\infty) = 0$.

By employing both the ionized and bound state wave functions,
one can compute the inner product appearing in the differential
cross section \geqn{dsigmadE} and subsequently derive the DM
production rate inside the Sun.

\begin{figure}[t]
\centering
\includegraphics[width=0.49\textwidth]{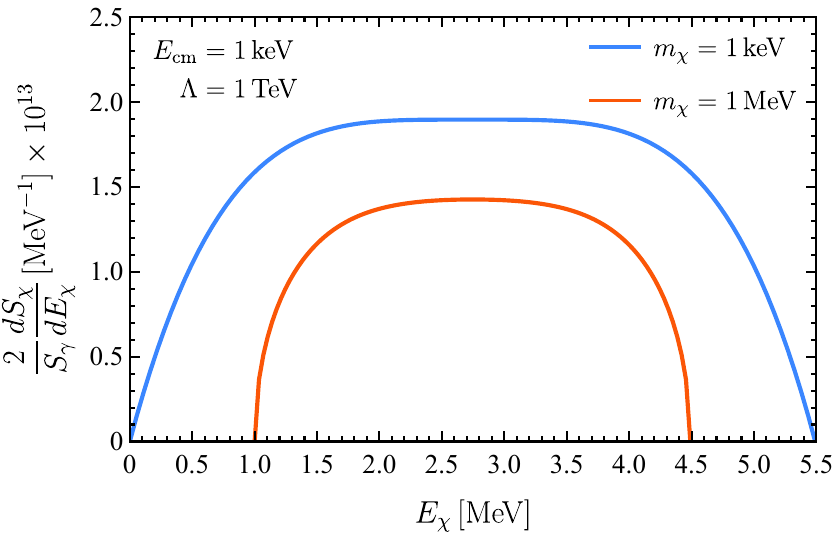}
\caption{The ratio of the DM and photon production
$S$-factors for the $p$-$d$ fusion.}
\label{fig:production}
\end{figure}

{\bf Production Rate} -- Since the deuteron number density is not publicly available
in the B16-GS98 solar model \cite{Vinyoles:2016djt}, it is more convenient
to deduce the DM production rate from the photon production
processes in the $p$-$d$ fusion. With exactly the same initial
states, their production rates in a volume
element $dV_\odot$ are proportional to their cross sections
($\sigma_\gamma$ for photon and $\sigma_\chi$ for DM),
\begin{equation}
    \frac{d^3 N_\chi}{dt dE_\chi dV_\odot} 
    = 
    2 \frac{1}{\left\langle \sigma_\gamma v^{pd}_{\mathrm{rel}} \right\rangle}
    \left\langle \frac{d \sigma_\chi  }{dE_\chi} v^{pd}_{\mathrm{rel}} \right\rangle
    \frac{d^2 N_\gamma}{dt dV_\odot},
    \label{production_rate_cross_section}
\end{equation}
where $\langle \cdots \rangle$ stands for thermal average \cite{Kolb:1990vq}
of the corresponding cross section times the relative
velocity $v^{pd}_{\mathrm{rel}}$.
The prefactor $2$ accounts for the two DM particles produced
in one fusion. Although the photon production rate
$d^2 N_\gamma / dt dV_\odot$ is not directly provided
in the Solar Model either, its value equals the sum of the $pp$
and $pep$ neutrino production rates
\cite{Vinyoles:2016djt}.
Thus, the DM production rate is linearly related to this sum throughout the Sun. Given that the $pp$
and $pep$ neutrino are primarily produced in the central region, DM is also predominantly produced in the solar core.

For small total kinetic energy $E_\mathrm{cm}$ of the initial-state nuclei,
the fusion is exponentially suppressed, since the incoming nuclei have to penetrate the Coulomb barrier \cite{Bertulani:2007bfy}.
We define the $S$-factor to accommodate the exponential dependence \cite{Bertulani:2007bfy},
$
    S_{\chi, \gamma} (E_\mathrm{cm})
    \equiv 
    \sigma_{\chi, \gamma}  (E_\mathrm{cm})
    E_\mathrm{cm}
    e^{2\pi \eta}
$
where $\eta \equiv Z_p Z_d \mu / (4\pi \hbar^2 k)$ is
a function of the proton and deuteron charge numbers
$Z_{p,d}$, the reduced mass
$\mu \equiv m_p m_d / (m_p + m_d)$, and the
center-of-mass momentum
$k = \sqrt{2 \mu E_\mathrm{cm}}$. 
Different from cross section, the $S$-factor tends to
be a constant at low energy,
$S_{\chi, \gamma} (E_\mathrm{cm}) \simeq
S_{\chi, \gamma}$ \cite{Bertulani:2007bfy}. 
Since $E_{\rm cm} e^{2 \pi \eta}$ is independent of
$E_\chi$, the cross sections $\sigma_{\chi, \gamma}$
in \geqn{production_rate_cross_section} can be
replaced by the $S$ factors,
\begin{equation}
    \frac{d^3 N_\chi}{dt dE_\chi dV_\odot} 
    \simeq 
    \frac{2}{S_\gamma} 
    \frac{d S_\chi}{dE_\chi} 
    \frac{d^2 N_\gamma}{dt dV_\odot} .
    \label{production_rate_S_factor}
\end{equation}
The ratio of production rates is proportional to the ratio of $S$-factors.
\gfig{fig:production} shows that the DM production rate
with $\Lambda = 1 \, \mathrm{TeV}$ is nearly 13 orders
of magnitude smaller than its photon counterpart. 
Although the cooling effect due to DM release is
negligible, the produced DM flux can be probed
at the DM direct detection experiments.

\section{Solar Attenuation with Three-Dimensional Analytic Boltzmann Equation Formalism}
\label{sec:attenuation}

When propagating inside the Sun, the DM particle scatters
with nuclei (mainly protons and $\alpha$ particles) and
roams until reaching the solar surface. These
scatterings would attenuate and soften the DM flux. 
Although the DM attenuation can be addressed with
both analytic
\cite{Starkman:1990nj,Kavanagh:2017cru,Bringmann:2018cvk,Xia:2020apm,Ge:2020yuf} and Monte Carlo
\cite{Collar:1992qc,Mahdawi:2017cxz,Emken:2021lgc,Xia:2021vbz,Chen:2021ifo,CDEX:2021cll}
methods, they have their own limitations. Especially,
the existing analytic methods based on the ballistic
approximation assume that DM propagates
in straight lines which is not appropriate for
multiple scatterings with large scattering angle.
For the convolutional approach that sums up all the
DM fluxes after multiple scattering
\cite{Cappiello:2023hza}, it currently only applies
to a homogeneous slab-shaped medium with isotropic
scattering.

We propose using the Boltzmann
method to precisely and efficiently calculate
the solar attenuation effect. The Boltzmann equation
describes the evolution of the distribution function
$f_\chi (\bm{r}_\chi, \bm{p}_\chi, t)$,
\begin{equation}
  \hat{\mathbf{L}} [f_\chi]
= 
  \boldsymbol{\mathrm{C}}_{\chi p}[f_\chi]
+ \boldsymbol{\mathrm{C}}_{\chi \alpha}[f_\chi]
+ \mathbf{C}_\mathrm{prod},
\label{Boltzmann_equation}
\end{equation}
where $\hat{\mathbf{L}} [f_\chi]$ is the Liouville
operator \cite{Kolb:1990vq,Lindquist:1966igj} and $\boldsymbol{\mathrm{C}}$s are collision terms
\cite{Kolb:1990vq,Du:2021jcj}.
Due to the spherical symmetry of the Sun, the distribution function $ f_\chi (\bm{r}_\chi, \bm{p}_\chi, t) $ is independent of the solid angles of $ \bm{r}_\chi $ and depends only on the radial coordinate $ r \equiv |\bm{r}_\chi| $. At any given location $ \bm{r}_\chi $, the momentum distribution of DM remains rotationally symmetric about the axis along $ \bm{r}_\chi $. Thus, $ f_\chi (\bm{r}_\chi, \bm{p}_\chi, t) $ is also independent of the azimuthal angle of $ \bm{p}_\chi $. Assuming the Sun has reached steady state, we can further neglect the time dependence in $ f_\chi (\bm{r}_\chi, \bm{p}_\chi, t) $. As a result, $ f_\chi (\bm{r}_\chi, \bm{p}_\chi, t) $ depends only on three variables: the radial distance $ r $, the angle between the position vector and the momentum (or its cosine $ u $), and the magnitude of the momentum (or energy $ E_\chi $), 
$f_\chi (\bm{r}_\chi, \bm{p}_\chi, t)
    \rightarrow 
    f_\chi (r, u, E_\chi)$.

Of the collision terms
\cite{Kolb:1990vq,Du:2021jcj},
the first two
$\boldsymbol{\mathrm{C}}_{\chi p}[f_\chi]$
and
$\boldsymbol{\mathrm{C}}_{\chi \alpha}[f_\chi]$
on the right-hand side
describe the elastic scattering of DM with a proton
or alpha particle target, respectively.
Each contains two contributions,
$\boldsymbol{\mathrm{C}}_{\chi p}[f_\chi] \equiv
\boldsymbol{\mathrm{C}}^{(1)}_{\chi p}[f_\chi]
+ \boldsymbol{\mathrm{C}}^{(2)}_{\chi p}[f_\chi]$ and
$\boldsymbol{\mathrm{C}}_{\chi \alpha}[f_\chi] \equiv
\boldsymbol{\mathrm{C}}^{(1)}_{\chi \alpha}[f_\chi]
+ \boldsymbol{\mathrm{C}}^{(2)}_{\chi \alpha}[f_\chi]$,
for flowing out or into the phase space point
under consideration \cite{Ge:2020yuf}.

The first $\chi$-$p$ scattering collision term,
$\mathbf{C}^{(1)}_{\chi p} [f_\chi]$, describes an
outflux of DM with kinematic variables $(u,E_\chi)$,
\begin{equation}
  \mathbf{C}^{(1)}_{\chi p} [f_\chi]
  \equiv 
- E_\chi f_\chi  
  \int \frac{g_p d^3 \bm{p}_p}{(2\pi)^3} 
  f_p (|\bm{p}_p|) 
  \sigma_{\chi p} v^{\chi p}_\mathrm{rel},
\label{collision_term_1}
\end{equation}
with $g_p = 2$ counting the proton spin.
The integral above is a thermal average of the
$\chi$-$p$ scattering cross section $\sigma_{\chi p}$
times the relative velocity $v^{\chi p}_\mathrm{rel}$
over the proton Boltzmann distribution
$f_p (|\bm{p}_p|)$. For comparison, the second term
$\boldsymbol{\mathrm{C}}_{\chi p}^{(2)}[f_\chi]$
describes a DM influx from the kinematic variables
$(u',E'_\chi)$,
\begin{align}
\hspace{-3mm}
  \mathbf{C}^{(2)}_{\chi p} [f_\chi] 
& \equiv 
  \int d\Pi_{p'} f_{p} (|\bm{p}'_p|) 
  \int \frac{ d\Omega'_{\chi}}{8 (2\pi)^2}
  f_{\chi} (r, u', E'_\chi) A
\label{collision_term_2}
\end{align}
with 
\begin{equation}
    A \equiv 
    \frac{ |\bm{p}'_{\chi}|^2 \overline{ | \mathcal{M}_{\chi p} |^2 } }
  {
  \left| |\bm{p}'_{\chi}| ( E_\chi - E_{p} ) - |\bm{p}_\chi - \bm{p}'_{p}| E'_{\chi} \cos \tilde{\theta} \right|
  },
\end{equation}
$d\Pi_{p'} \equiv \frac{g_p \, d^3 \bm{p}'_{p}}{(2\pi)^3 2E_p}$, and $\overline{ |\mathcal{M}|^2_{\chi p} } = 4 m_p^2 / \Lambda^2$ being the spin-averaged squared $\chi$-$p$ scattering
amplitude.
The solid angle $\Omega'_{\chi}$ is for the incoming
DM particle while $\tilde{\theta}$ is the angle between the initial-state DM momentum $\bm{p}'_{\chi}$ and
the difference $(\bm{p}_\chi - \bm{p}'_{p})$
between the final DM ($\bm p_\chi$) and
the initial proton ($\bm p'_p$) momenta.
Note that the incoming DM energy $E'_{\chi}$ is not
an independent variable here but is determined by
the energy-momentum conservation.

The two integration terms \geqn{collision_term_1}
and \geqn{collision_term_2} are complicated. 
Since the proton mass ($\simeq \mathrm{GeV}$) is
much larger than the proton and DM momentum as
well as the DM energy ($\sim \mathrm{MeV}$),
the two collision terms can be expanded up to
$\mathcal{O} \left( 1/m_p \right)$ for
convenience.
The expansion of $\sigma_{\chi p} v^{\chi p}_\mathrm{rel}$ gives,
\begin{equation}
    \sigma_{\chi p} v^{\chi p}_\mathrm{rel}
    \simeq
    \sigma_{\chi p}^{\mathrm{LO}}
    v_\chi 
    \left(
      1 
      - 
      \frac{ 2 E_\chi }{m_p} 
      -
      \frac{ E_\chi |\bm{p}_p|}{ m_p |\bm{p}_\chi| } 
      \cos \theta_{\chi p}
    \right),
\end{equation}
where $ \sigma_{\chi p}^{\mathrm{LO}} \equiv 1 / 4\pi \Lambda^2 $ is the leading-order contribution to the $\chi$-$ p $ scattering cross section. 
The term proportional to $\cos \theta_{\chi p}$ vanishes after integrating over the solid angles of $\bm{p}_p$.
The first collision term then becomes, 
\begin{equation}
    \boldsymbol{\mathrm{C}}_{\chi p}^{(1)}[f_\chi]
    \simeq
    - |\bm{p}_\chi| 
    n_p \sigma_{\chi p}^{\mathrm{LO}} 
    f_\chi (r, u, E_{\chi}) 
    \left(
    1 
    -  \frac{2 E_\chi}{m_p}
    \right),
\end{equation} 
where $ n_p $ is the proton number density.

\begin{figure}[t]
    \centering
    \includegraphics[width=5.5cm]{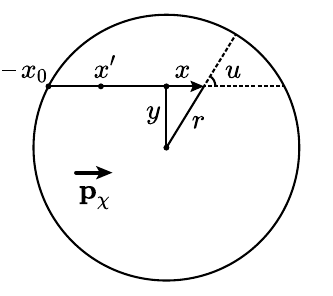}
    \caption{Illustration of the geometric meaning of $x$ and $y$. $x'$ denotes the integration variable in the integral equation form of the Boltzmann equation. $-x_0$ is on the solar surface.}
    \label{fig:ru_xy_illustration}
\end{figure}

In contrast, the expansion of \geqn{collision_term_2} is more intricate due to the dependence of $ f_\chi (r, u', E'_\chi) $ on the variable $ E'_\chi $, which is a function of integration variables and should also be expanded. To address this, we perform a Taylor expansion of the unknown function $ f_\chi (r, u', E'_\chi) $ around $ E'_\chi = E_\chi $,
\begin{equation}
  f_\chi
\approx 
  f_\chi(r, u', E_{\chi})
+ \frac {\partial f_\chi(r, u', E_{\chi})}
        {\partial E_\chi}
  (E'_{\chi} - E_\chi).
\label{eq:expand_f}
\end{equation}
This allows extraction of $ E'_\chi $ from the function $ f_\chi $, enabling further expansion in $ \mathcal{O} \left( 1/m_p \right) $ and subsequent integration as well as simplification. 
Since $f_\chi(r, u', E_\chi)$ in both terms
depends only on $u'$ but no longer $E'_\chi$, 
the integrations over $d\Pi_{p'}$ and the azimuthal
angle $d\phi'$ of $\Omega'_\chi$ only applies to
$E'_\chi$. Defining an averaged $\bar{E}'_{\chi}$,
\begin{equation}
  \bar{E}'_{\chi}
  \equiv 
  \frac{
    \int d \phi' d\Pi_{p'} f_{p} (|\bm{p}'_{p}|)  
    A
    E'_{\chi}
  }{
    \int d \phi' d\Pi_{p'} f_{p} (|\bm{p}'_{p}|)  
    A
  },
  \label{Ebar_definition}
\end{equation}
one can factor out the term
$\int d\phi'\, d\Pi_{p'}\, f_{p}(|\bm{p}'_{p}|)\, A$
from the integral in \geqn{collision_term_2} and
simply replace $E'_{\chi}$ in \geqn{eq:expand_f}
with $\bar{E}'_{\chi}$. Then, inserting
\geqn{eq:expand_f} yields an integral expression
for the function $f_\chi(r,u',\bar{E}'_{\chi})$,
\begin{equation}
\hspace{-3mm}
  \boldsymbol{\mathrm{C}}_{\chi p}^{(2)}[f_\chi] 
\simeq
  \int \frac{ d u'}{32 \pi^2} 
  f_\chi (r, u', \bar{E}'_{\chi})
  \int d \phi' d\Pi_{p'} f_{p} (|\bm{p}'_{p}|)  
  A ,
\label{second_collision_term_form3.5}
\end{equation} 
The factor $A$ defined in \geqn{collision_term_2}
can be expanded up to $\mathcal{O}(1/m_p)$ as
\begin{equation}
  A 
\simeq  
  \frac{ |\bm{p}_\chi| }{ m_p }
  \overline{ | \mathcal{M}_{\chi p} |^2 }
\left[
  1 
+  \frac{ 2 E_\chi }{m_p} ( 1 - \cos \theta_{\chi \chi'} )
+ \cdots 
\right],
\label{Aexpansion}
\end{equation}
where $\theta_{\chi \chi'}$ denotes the opening
angle between the initial- and final-state DM
particles. The omitted terms $\cdots$ in \geqn{Aexpansion}
contains $\cos \theta_{\chi' p'}$ and $\cos \theta_{\chi p'}$,
which vanish after integration over the solid angle
of $\bm{p}_{p'}$.
Besides, the integration of 
$\cos \theta_{\chi \chi'} = u u' + \sqrt{1-u^2}\sqrt{1-u^{\prime 2}} \cos \phi'$ 
over $d\phi'$ yields $2 \pi u u'$ to give,
\begin{equation}
\hspace{-3mm}
  \int d \phi' d\Pi_{p'} f_{p}   A
\simeq 
  16 \pi^2 |\bm{p}_\chi| n_p \sigma_{\chi p}^{\mathrm{LO}}
\left[
  1 
+ \frac{ 2 E_\chi }{m_p } ( 1 - u u' )
\right].
\label{eqtemp}
\end{equation}

Similarly, the factor $\bar{E}'_{\chi}$ in Eq.~\eqref{Ebar_definition} can be expanded up to $\mathcal{O}(1/m_p)$.
Using the expansion of $E'_\chi$, 
\begin{equation}
    \begin{aligned}
        E'_{\chi}
          ={}
          &E_\chi
          +
          \frac{|\bm{p}_\chi|^2}{m_p} ( 1 - \cos \theta_{\chi \chi'}) + \cdots, 
    \end{aligned}
\end{equation}
the integration in Eq.~\eqref{Ebar_definition} can be evaluated in the same manner as in Eq.~\eqref{eqtemp}, giving
\begin{equation}
    \bar{E}'_{\chi} \simeq E_\chi + \frac{|\bm{p}_\chi|^2}{m_p} ( 1 - u u') .
\end{equation}

Putting things together, the final expansion of
\geqn{collision_term_2} is given by  
\begin{align}
  \boldsymbol{\mathrm{C}}_{\chi p}^{(2)}[f_\chi] 
& \simeq
  |\bm{p}_\chi|
  n_p \sigma_{\chi p}^{\mathrm{LO}} 
  \int \frac{ d u'}{2} 
  f_\chi (r, u', \bar{E}'_{\chi}) 
\nonumber
\\
& \hspace{2.4cm} \times
\left[
  1 
+ \frac{ 2 E_\chi }{m_p } ( 1 - u u' )
\right] .
\end{align}
The integrals are now greatly simplified.

Similar simplification can also apply
to the $\chi$-$\alpha$ collision terms,
$\boldsymbol{\mathrm{C}}_{\chi \alpha}^{(1,2)}[f_\chi]$.
Note that the $\chi$-$\alpha$ scattering cross section
$\sigma_{\chi \alpha}^{\mathrm{LO}} = Z_\alpha^2
\sigma_{\chi p}^{\mathrm{LO}}$
is coherently enhanced by the ${}^4 \mathrm{He}$
charge $Z_\alpha = 2$.

The remaining $\mathbf{C}_{\rm prod}
    \equiv 
    \frac{2\pi^2}{ |\bm{p}_\chi| }
    \frac{d^3 N_\chi}{dt dE_\chi dV_\odot}$
is actually a source term from the DM production
that also happens all over the Sun as given in
\geqn{production_rate_S_factor}.

To uniquely solve the differential Boltzmann equation
in \geqn{Boltzmann_equation}, we need a boundary
condition that no DM particle enters the solar surface,
$f_\chi(R_\odot, u, E_\chi) = 0$ for $u \leq 0$,
where $R_\odot$ is the solar radius. 
In the spherical coordinate system,
the Liouville operator appearing on the left-hand side of \geqn{Boltzmann_equation} can be expressed as,
\begin{equation}
  \hat{\boldsymbol{\mathrm{L}}}[f_\chi] = |\bm{p}_\chi| \left( u \frac{\partial f_\chi}{\partial r} + \frac{1-u^2}{r} \frac{\partial f_\chi}{\partial u} \right) .
  \label{Liouville_operator_final}
\end{equation}
By making the variable substitution $x \equiv r u$
and $y \equiv r \sqrt{1-u^2}$, the Liouville operator
reduces to a single derivative,
$\hat {\bf L}[f_\chi] = |{\bm p}_\chi| \partial_x f_\chi$.
The variable $x $ denotes the DM position along its propagation line, and together with $y$, they form a Cartesian coordinate system (see in \gfig{fig:ru_xy_illustration}).
The solution to the Boltzmann equation 
\geqn{Boltzmann_equation} is then an integral equation
with the integration constant fixed by the boundary
condition, 
\begin{widetext}
    \begin{align}
        \label{integral_function}
          f_\chi(r,u,E_\chi) 
          &= 
          \int_{-x_0}^{x} dx' \, 
          e^{-\int^{x}_{x'} \sum_i n_i(r(x'',y)) \sigma'_{\chi i}  dx''} 
          \left[
            \frac{2\pi^2}{|\bm{p}_\chi|^2 } 
            \frac{d^3 N_\chi}{dt dE_\chi dV_\odot} (r(x',y), E_\chi)
          \right] \\
          &+ 
          \int_{-x_0}^{x} dx' \, 
          e^{-\int^x_{x'} \sum_i n_i(r(x'',y)) \sigma'_{\chi i}  dx''}
          \sum_i 
          n_i (r(x',y))  \sigma^\mathrm{LO}_{\chi i} 
          \int \frac{du'}{2}
          f_\chi(r(x',y),u',\bar{E}'_i)
          \left[ 1 + \frac{2 E_\chi ( 1 - u u' ) }{m_i} \right] , \nonumber
    \end{align}
\end{widetext} 
where $i = (p, \alpha)$ labels the scattering targets.
In addition,
$\sigma'_{\chi i} \equiv \sigma^\mathrm{LO}_{\chi i} \left( 1 - 2 E_\chi / m_i \right)$
and $ \bar{E}'_{i} \equiv E_\chi + |\bm{p}_\chi|^2 ( 1 - u u')/m_i $.
The radial function is defined as $r(x,y) \equiv \sqrt{x^2 + y^2}$
and $x_0 \equiv \sqrt{R_\odot^2-y^2}$ corresponds to the solar surface. 

This integral equation has clear physical interpretation. 
The first term on the right-hand side represents the contribution to $f_\chi(r,u,E_\chi)$ from the DM production at position $(x',y)$ that propagates from $x'$ to $x$ without scattering. The exponential factor corresponds to the probability of propagation without scattering.
The second term accounts for the contribution to $f_\chi(r,u,E_\chi)$ from those DM particles, originally characterized by the kinematic variables $(u',\bar{E}'_i)$, that scatter at position $(x',y)$ into $(u,E_\chi)$ and then propagate from $x'$ to $x$ without further scattering.%

\begin{figure}[t]
\centering
\includegraphics[width=0.48\textwidth]{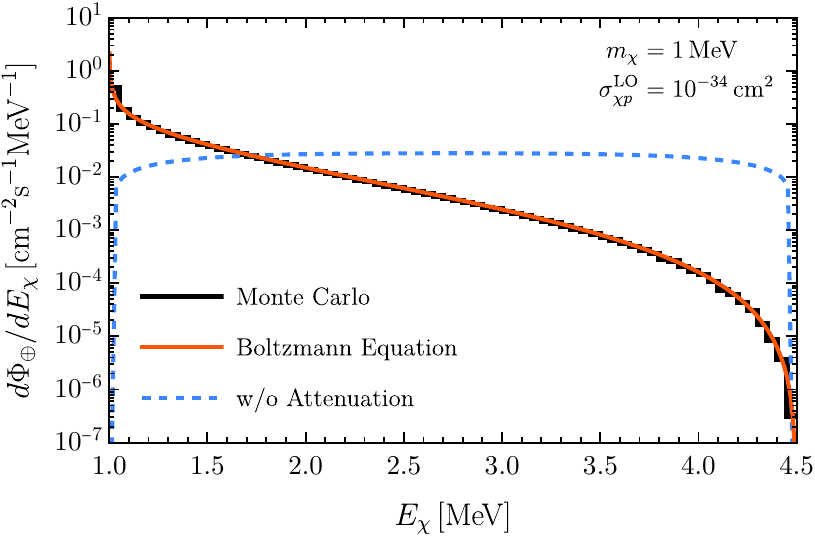}
\caption{The solar DM flux spectrum arriving at Earth for
$m_\chi = 1\, \mathrm{MeV}$ and
$\sigma_{\chi p}^{\mathrm{LO}} = 10^{-34} \, \mathrm{cm}^2$.
The red curve is from the analytic Boltzmann equation method
while the black curve is from Monte Carlo simulation based on
\texttt{DarkProp} \cite{DarkProp:v0.3}. The blue dashed curve
is obtained assuming no attenuation.}
\label{fig:attenuation}
\end{figure}

When propagating
from the solar surface to our Earth, the DM flux
\begin{equation}
    \frac{d\Phi_\oplus}{dE_\chi} 
    =
    \frac{R_\odot^2}{\mathrm{AU}^2}
    |\bm{p}_\chi|^2 
    \int_0^1
    \frac {du}{4 \pi^2}
    u f (R_\odot, u, E_\chi),
\end{equation}
is diluted by a factor of
$R_\odot^2 / \mathrm{AU}^2$ where 
$\mathrm{AU}$ is the astronomical unit.
\gfig{fig:attenuation} shows the DM flux spectrum
at the Earth for $m_\chi = 1\, \mathrm{MeV}$ and
$\sigma_{\chi p}^{\mathrm{LO}} = 10^{-34} \, \mathrm{cm}^2$.
Our result (red thin line)
is verified by the Monte Carlo simulation
with \texttt{DarkProp} \cite{DarkProp:v0.3}. 
The solar attenuation effect can significantly
change the DM spectrum. The quite flat spectrum
in the middle as shown by \gfig{fig:production} is
attenuated to a low energy peak in
\gfig{fig:attenuation}.

We note that there is a boundary in scattering cross section between severe and mild attenuation.
It can be estimated using the number of scatterings of DM across the sun,
$
    \int_0^{R_\odot} ( n_p \sigma_{\chi p}^{\mathrm{LO}} + n_\alpha \sigma_{\chi \alpha}^{\mathrm{LO}} ) dr \sim 30.
$
We choose the number of scatterings to be 30 typically instead of 1 because DM loses only $\sim 10^{-3}$ of its energy in each scattering and it can scatter several times before escaping without losing too much energy. 
From this we can derive the typical boundary as $\sigma_{\chi p}^{\mathrm{LO}} \sim 4 \times 10^{-35} \, \mathrm{cm}^2$. 
For smaller scattering cross section, the attenuation effects can be neglected.
%

\section{Direct Detection of MeV Solar Dark Matter}
\label{sec:detection}

When reaching Earth, the solar DM can be detected in direct detection experiments. 
For xenon-based detectors, the maximal recoil energy for a xenon nucleus with mass $m_{\mathrm{Xe}}$,
$T_\mathrm{N}^\mathrm{Xe} \simeq 2 |\bm{p}_\chi|^2 / m_{\mathrm{Xe}} \simeq 0.4 \, \mathrm{keV}$
where $|\bm{p}_\chi| \simeq 5.5 \, \mathrm{MeV}$
is the maximal momentum as shown in \gfig{fig:production},
is below the Xenon1T (0.7\,keV) and PandaX-4T (0.77\,keV) S2-only thresholds \cite{XENON:2019gfn,PandaX:2022xqx}.
For argon-based detectors,
the recoil energy can reach
$T_\mathrm{N}^\mathrm{Ar} \simeq 1.5 \, \mathrm{keV}$
to exceed the threshold (0.6\,keV, corresponding to
the number of ionization electrons $N_{e^-} = 4$)
of DarkSide-50 \cite{DarkSide-50:2022qzh,DarkSide:2022dhx}.

In the limit of weak $\chi$-$p$ coupling, the solar DM
production and event rate decreases accordingly. In the
strong coupling limit, although the solar DM can be
abundantly produced, the solar attenuation effect becomes
severe and DM loses too much energy inside the Sun such
that the DM event rate above threshold also decreases.
Furthermore, the DM detection spectrum drops at large
recoil energy as shown in \gfig{fig:attenuation}. 
Thus, there is a maximum of the solar DM event rate, $\simeq 10^{-5} / N_{e^-} \cdot \mathrm{kg} \cdot \mathrm{day}$, at the threshold $N_{e^-} = 4$.
For comparison, the background of Darkside-50 at $N_{e^-} = 4$ is around $10^{-2} / N_{e^-} \cdot \mathrm{kg} \cdot \mathrm{day}$ \cite{DarkSide-50:2022qzh}.
Therefore, the sensitivity of Darkside-50 is not sufficient to detect solar DM.
However, the next-generation detector, DarkSide-LowMass
(DS-LM) \cite{GlobalArgonDarkMatter:2022ppc}, with larger
fiducial mass ($\simeq 1 \, \mathrm{ton}$ compared to
$\simeq 20 \, \mathrm{kg}$ in DarkSide-50), lower
threshold, and reduced background ($\simeq 10^{-4} / N_{e^-} \cdot \mathrm{kg} \cdot \mathrm{day}$
at the threshold $N_{e^-} = 2$), is capable of
detecting solar DM. At DarkSide-LowMass, the major
background from ${}^{39} \mathrm{Ar}$ is projected to be
$7.3\,\mu \mathrm{Bq/kg}$ or $73\,\mu \mathrm{Bq/kg}$~\cite{GlobalArgonDarkMatter:2022ppc}.

\begin{figure}[t]
\centering
\includegraphics[width=0.48\textwidth]{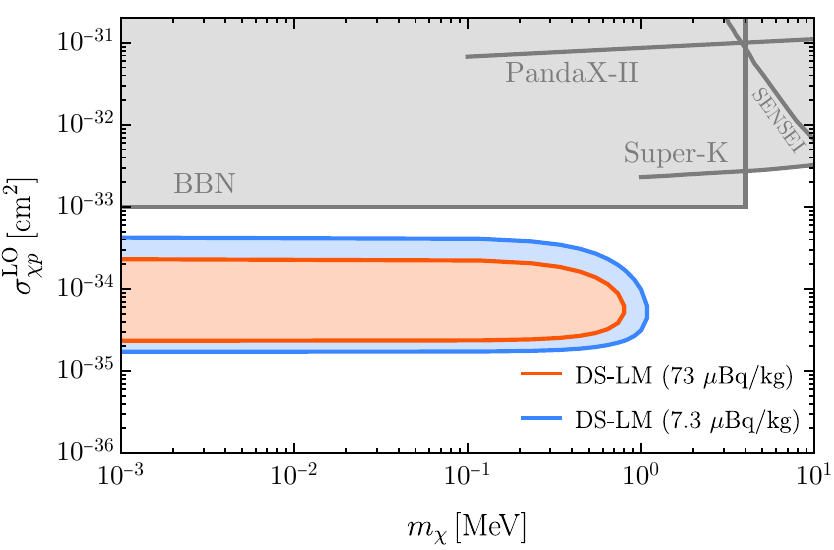}
\caption{The projected exclusion regions for the solar DM parameter
space at DarkSide-LowMass with a ${}^{39} \mathrm{Ar}$ background
level of $73\,\mu \mathrm{Bq/kg}$ (red) or
$7.3\,\mu \mathrm{Bq/kg}$ (blue). For comparison, the
exclusion limits from 
PandaX \cite{PandaX-II:2021kai}
and Super-Kamiokande \cite{Super-Kamiokande:2022ncz} on the
cosmic-ray boosted DM,
SENSEI \cite{SENSEI:2023zdf}, 
as well as the BBN constraint are also
shown. }
\label{fig:detection}
\end{figure}

Assuming a $1 \, \mathrm{ton}\cdot \mathrm{year}$ exposure, we show the projected
$90\%\,\mathrm{C.L.}$ limits as colored curves in \gfig{fig:detection}.
The DarkSide-LowMass experiment
is sensitive to the sub-MeV solar DM with 
a scattering cross section
$10^{-35}~\mathrm{cm}^2 \lesssim \sigma_{\chi p}^{\mathrm{LO}} \lesssim 4\times10^{-34}~\mathrm{cm}^2$,
which is two orders lower than the current limits from the cosmic-ray boosted DM
\cite{PandaX-II:2021kai,Super-Kamiokande:2022ncz},
while the conventional direct detection can
only reach $10^{-27} \, \mathrm{cm}^2$
at the SENSEI experiment \cite{SENSEI:2023zdf}.
Since the relevant DM mass range is well below the production energy, 
both the upper and lower boundaries are almost independent
of the DM mass and can extend to very tiny mass.

Usually, sub-MeV DM is stringently constrained by the
big bang nucleosynthesis (BBN)
\cite{Berezhiani:2012ru,Nollett:2013pwa,Green:2017ybv,Krnjaic:2019dzc,Sabti:2019mhn,An:2022sva}.
If thermally coupled to the SM plasma,
a complex scalar DM with mass $\lesssim 4\,\mathrm{MeV}$ is 
excluded by BBN \cite{Sabti:2019mhn}. 
Such constraint can be alleviated if DM decouples with SM
particles first and then is diluted to a smaller density
before BBN \cite{Evans:2019jcs}.
The dilution can be induced by a heavy out-of-equilibrium
particle decaying into SM particles.
Note that the decay process may also dilute neutrinos. 
To keep neutrinos unchanged, it should happen earlier
than neutrino decoupling, which requires the decay width
$\Gamma > 10^{-23} \, \mathrm{GeV}$
\cite{deSalas:2015glj}.
After dilution, the complex scalar DM should have a lower
temperature $T_\chi$ than that of the SM plasma
$T_\mathrm{SM}$ , $T_\chi < 0.77 \, T_\mathrm{SM}$
\cite{Yeh:2022heq}, to be compatible with BBN.
Equivalently, this requires that DM decouples
earlier than the heavy particle decay,
$H(T_\mathrm{dec}) > 2.25 \, \Gamma$ where
$H(T_\mathrm{dec})$ denotes the Hubble rate at the
decoupling temperature $T_\mathrm{dec}$.
Thus, the minimal decoupling temperature is
$T_\mathrm{dec} > 7.05 \, \mathrm{MeV}$.
In our scenario, the DM decoupling is controlled
by the tree level $p + \bar{p} \to \chi + \chi^*$, $\gamma + p \to p + \chi + \chi^*$
process, as well as the loop-induced
$\gamma + \gamma \to \chi + \chi^*$ process. 
Since the proton number density is exponentially
suppressed at low temperature, the latter process
dominates
and the aforementioned lower bound on $T_{\rm dec}$
then transfers to $\Lambda > 176 \, \mathrm{GeV}$.
In other words, a light complex scalar DM can be
compatible with BBN, if the $\chi$-$p$ scattering
cross section is small enough,
$\sigma_{\chi p}^{\mathrm{LO}} < 9.97 \times 10^{-34} \, \mathrm{cm}^2$,  
as shown in \gfig{fig:detection}.

\section{Conclusion} \label{sec:conclusion}

Not just the thermal and atomic processes inside the Sun
can evaporate DM or produce light DM such as axion, but
also the nuclear fusion can produce energetic MeV DM
particles. We provide a concrete example of the proton
deuteron fusion process that during the $p$-$d$ system
transition from an ionized state to its bound state,
namely the ${}^3$He nuclei, a pair of DM particles are
produced. With an energy release of 5.5\,MeV, the
produced DM can overcome the direct detection threshold.
Being not strongly constrained, the produced solar DM
can experience strong attenuation inside the Sun. With
spherical symmetry, the Boltzmann equation can be used
to describe the attenuation quite well.

\section*{Acknowledgements}

The authors would like to thank Junting Huang and
Yi Wang for useful discussions.
Chuan-Yang Xing and Chen Xia are supported by the
National Natural Science Foundation of China (Nos. 12247141,
12247148). Shao-Feng Ge is supported by the National Natural
Science Foundation of China (Nos. 12375101, 12425506, 12090060, 12090064)
and the SJTU Double First Class start-up fund (WF220442604).
SFG is also an affiliate member of Kavli IPMU, University of Tokyo.

\bibliographystyle{utphysGe}
\bibliography{solarDM}

\end{document}